\documentstyle [epsfig,wrapfig,12pt]{article}
\begin{document}
\hfill  ORNL-CCIP-94-14
\vspace{1.0cm}
{
\begin{center}
{\Large\bf Hadronic Molecules and Low-Energy\\
Hadron-Hadron Scattering Amplitudes}
\footnote[1]{This is an expanded version of results presented at
Few Body XIV, Williamsburg, Virginia (26-31 May
1994).}\\
\vspace{1cm}
T.Barnes\\
Physics Division and Center for Computationally Intensive Physics\\
Oak Ridge National Laboratory, Oak Ridge, TN 37831\\
and\\
Department of Physics and Astronomy\\
University of Tennessee\\
Knoxville, TN 37996\\
\date{}
\end{center}
}

\begin{abstract}
Recently there has been considerable interest in the subject of molecules,
which are weakly bound states of hadron pairs. The question of the existence of
molecules is closely related to the more general problem of the determination
of low energy hadron-hadron scattering amplitudes, which is widely believed to
require nonperturbative methods. In this contribution we report on quark model
calculations using a simple perturbative scattering mechanism, one gluon
exchange followed by constituent interchange. We refer to the associated
diagrams as ``quark Born diagrams". For the cases chosen to isolate this
mechanism, I=2 $\pi\pi$, I=3/2 K$\pi$ and KN the results are usually in good
agreement with experimental S-wave scattering amplitudes given standard
potential model parameters, and for NN we find perturbative results very
similar to the nonperturbative hard cores of Oka and Yazaki. We also discuss
our findings for other less familiar channels; these include predictions of
vector-vector bound states, one of which may be the $\theta(1710)$.

\end{abstract}

\section{Introduction}
\thispagestyle{empty}

Nuclear physics has provided us with a very large number of examples
of weakly bound states of
largely undistorted hadrons. Since this system is actually rather unfavorable
for the formation of bound states due to the repulsive core interaction,
one is naturally lead to enquire into the possibility that other nuclear
species exist which have constituents other than nucleons. This type of state,
consisting of two or more weakly bound hadrons, is referred to as a
``hadronic molecule", and many candidates for these states are now known.
(For a recent review see Ref.\cite{molrev}, which gives extensive references
to the literature.)

The search for molecules is largely driven by experiment at present, because
hadronic forces are rather complicated and difficult to model accurately
without experimental guidance. Clearly the subject would be aided
by a better theoretical understanding of hadron-hadron interactions, so
one can predict
attractive
channels
in which molecules might form. This description should involve QCD degrees
of freedom (quarks and gluon) directly, so that it can be applied to
scattering of excited
hadrons as well as ground states, which are more usually
modelled using effective
lagrangians.

Hadron-hadron interactions are complicated in part due to the presence of
several different scattering mechanisms.
The important
contributions we identify are t-channel pion exchange
(heavier mesons give very short ranged forces, so they are unlikely to
be important in t-channel \cite{novec}),
s-channel resonance production (very important
when allowed) and quark interchange. By a careful choice of channel we
can largely eliminate the first two effects, and study the short range
quark exchange forces in isolation. We have had considerable success in
describing this contribution using a simple OGE and constituent interchange
model, which leads to analytic results for scattering amplitudes given
SHO wavefunctions. Our method
(introduced in Ref.\cite{Swan,BS})
reduces the calculation
of hadron scattering amplitudes in the nonrelativistic quark model to a
straightforward diagrammatic technique,
and some of our results are summarized below.

\vskip 0.5cm
\section{Method}

We assume that nonresonant hadron scattering is dominated by the lowest-order
perturbative QCD process, specifically one-gluon-exchange followed by
constituent (quark) interchange. Nonperturbative QCD is assumed to contribute
to scattering only through the asymptotic hadron wavefunctions
(or in the most detailed study as a single scattering interaction through a
linear confining potential \cite{Swan}).
We use a conventional nonrelativistic quark-model Hamiltonian of the form
\begin{equation}
H = \sum_{i=1}^4 \left( {p_i^2 \over 2 m_i} + m_i \right) + \sum_{i<j}
\left[ V_{conf}(r_{ij}) + V_{hyp}(r_{ij})\; \vec S_i \cdot \vec S_j
\; \right] (\lambda^a_i/2) \cdot (\lambda^a_j/2) \ ,
\end {equation}
\noindent
where
$V_{hyp} = -(8 \pi \alpha_s / 3 m_i m_j)\; \delta(\vec r_{ij})$
is the contact color-hyperfine interaction and
$V_{conf}$ is the spin-independent
confining potential. We then calculate the Born-order scattering amplitude
and corresponding phase shifts, using relativistic phase space and
kinematics for the external hadrons.
In most cases discussed here the hyperfine
term dominates \cite{Swan}, and we show explicit results for this interaction
and use simple Gaussian hadron wavefunctions for illustration unless stated
otherwise.

\vskip 0.5cm
In a two-meson scattering process without identical quarks the OGE + CI
scattering mechanism leads to four scattering diagrams, since there are $2
!\cdot 2 ! $ permutations of OGE interactions between the
incoming lines. One such diagram is shown below, and the rules for
generating and evaluating these diagrams are given in reference \cite{BS}.

\vskip 1cm

\setlength{\unitlength}{1.5pt}
\begin{picture}(180,70)(-50,-5)
\put(30,30) {\makebox(0,0)[1]{{\it capture}$_1 \ \ $  = } }
\put(75,50) {\makebox(0,0)[1]{A} }
\put(75,10) {\makebox(0,0)[1]{B} }
\put(160,10) {\makebox(0,0)[1]{D.} }
\put(160,50) {\makebox(0,0)[1]{C} }
\put(76,0){
\begin{picture}(75,60)(-2,0)
\multiput(5,15)(0,40){2}{\multiput(0,0)(55,0){2}{\vector(1,0){5}}}
\multiput(15,5)(0,40){2}{\multiput(0,0)(55,0){2}{\vector(-1,0){5}}}
\multiput(5,5)(60,10){2}{\multiput(0,0)(0,40){2}{\line(1,0){5}}}
\multiput(10,15)(0,40){2}{\line(1,0){50}}
\multiput(15,5)(0,40){2}{\multiput(0,0)(35,0){2}{\line(1,0){15}}}
\put(30,5){\line(1,2){20}}
\put(30,45){\line(1,-2){20}}
\put(20,5){\dashbox{2}(0,50){}}
\multiput(20,5)(0,50){2}{\circle*{2}}
\end{picture}
}
\end{picture}

\eject
\section{Results for $\pi\pi$, K$\pi$, KN and NN}

\vskip 0.5cm
\noindent
{\it a) I=2 $\pi\pi$}
\vskip 0.5cm

\begin{wrapfigure}{r}{4.0in}
\epsfig{file=h1.epsm,width=4.0in}
\caption{I=2 $\pi\pi$ S-wave phase shift.}
\label{fig 1}
\end{wrapfigure}
The I=2 $\pi\pi$ scattering amplitude due to the contact spin-spin interaction
with Gaussian wavefunctions may be evaluated in closed form
\cite{BS}. The S-wave is
dominant at energies studied experimentally, and is given by
(2) below.
The two free parameters
are reasonably well established in quark model
phenomenology, $\alpha_s/m_q^2 \approx (0.6/0.33^2)$ GeV$^{-2}$ and $\beta_\pi
\approx 0.3 $ GeV. We fix $\alpha_s/m_q^2$ at this standard value and fit the
less well-determined Gaussian width parameter $\beta_\pi$ to the data of
Hoogland {\it et al.} \cite{Hoog}. This is shown in figure 1, and the
fitted value is $\beta_\pi=0.337$ GeV, close to expectations. There is a clear
need for an experimental determination of the phase shift above $M_{\pi\pi}=
1.5$ GeV, which we predict to be near an extremum.

\vskip 0.5cm
The S-wave phase shift is
\begin{equation}
\sin \delta_2^{(0)}
= - \Bigg\{
{4\alpha_s \beta_\pi^2 \over 9 m_q^2 }
\;
\Bigg( 1 - e^{-(s-4M_\pi^2)/8\beta_\pi^2 }
+{(s-4M_\pi^2)\over \sqrt{27}\beta_\pi^2  }\,
e^{-(s-4M_\pi^2)/12\beta_\pi^2  }  \,
\Bigg) /
\sqrt{ 1-4M_\pi^2/s}
\Bigg\}
\ ,
\end{equation}
which leads to an
I=2 $\pi\pi$ scattering length of
\begin{equation}
a_2^{(\pi\pi)} = -{2\over 9} \bigg(1 + {8\over \sqrt{27}} \bigg)
{\alpha_s M_\pi \over m_q^2} \ .
\end{equation}
This is numerically
$a_2^{(\pi\pi)} /  M_\pi^{-1}   =  -0.059$,
in good agreement with Weinstein's (unscaled) variational result
\cite{JW}, and with the PCAC result of Weinberg \cite{Wei} and the
recent parametrization of Morgan and Pennington \cite{MP}; the latter
two references both find
$a_2^{(\pi\pi)} / M_\pi^{-1} = -0.06 $.
\vskip 0.5cm
\noindent
{\it b) I=3/2 } K$\pi$
\vskip 0.5cm

Another annihilation-free channel is I=3/2 K$\pi$
\cite{BSW}. Here we have four free parameters,
$\alpha_s/m_q^2$, $m_q/m_s$, $\beta_\pi$ and $\beta_K$. We leave
$\alpha_s/m_q^2$ and $\beta_\pi$ equal to their $\pi\pi$ values. The
phase shifts are found to be rather insensitive to the relative kaon/pion
length scale, so we set $\beta_K=\beta_\pi$; this leaves only
$m_q/m_s$ undetermined, and typical constituent quark parameters suggest
$m_q/m_s\approx 0.33$ GeV/0.55 GeV = 0.6.

\vskip 0.5cm

When we use the S-wave phase shift
data of
Estabrooks {\it et al.} \cite{Esta} to determine $m_s$
we find that the optimum ratio
is $m_q/m_s=0.677$, similar to expectations.

\begin{wrapfigure}{r}{4.0in}
\epsfig{file=h2.epsm,width=4.0in}
\caption{I=3/2 K$\pi$ S-wave and P-wave phase shifts.}
\label{fig 2}
\end{wrapfigure}

Imposing $m_q\neq m_s$ has an interesting effect; it induces a P-wave
amplitude.
The theoretical S-wave and P-wave phase shifts are shown in figure 2. Note that
the previous prediction of an extremum in the I=2 $\pi\pi$ S-wave phase shift
near 1.5 GeV (in figure 1) also holds for its SU(3)-partner I=3/2 K$\pi$ S-wave
phase shift (figure 2). I=3/2 K$\pi$ experimental data are available
at higher invariant mass than I=2 $\pi\pi$,
and appear to support this prediction. Our result for
the scattering length,

\begin{equation}
a_{3/2}^{(K\pi)} =   - { 2\alpha_s \over 9 m_q^2 (M_\pi^{-1} + M_K^{-1}) }
\left[ 1 +
  \left( {4 \beta_\pi^2 \over  2\beta_\pi^2 + \beta_K^2 }\right)^{3/2}
+  {m_q\over m_s} \Bigg\{ \left( {4 \beta_K^2  \over 2\beta_K^2  +\beta_\pi^2 }
\right)^{3/2} +
\left( {2 \beta_K\beta_\pi \over \beta_K^2  +\beta_\pi^2 }\right)^3
\Bigg\} \right]
\end{equation}
corresponds to about $-0.077/M_\pi$
with our preferred parameter set; this is
consistent with the PCAC
result of $\approx -0.07/M_\pi$ \cite{Wein} and with the (rather
wide) range of reported experimental values. Finally, the predicted and
observed P-waves are qualitatively consistent.

\vskip 0.2cm
\noindent
{\it c) }KN
\vskip 0.2cm

\begin{wrapfigure}{r}{4.0in}
\epsfig{file=h3.epsm,width=4.0in}
\caption{KN S-wave phase shifts; I=0 (open points) and I=1 (solid points).}
\label{fig 3}
\end{wrapfigure}
One may also study meson-baryon and baryon-baryon scattering using quark Born
diagrams. We have considered KN scattering in detail \cite{BSKN}, since there
are no valence annihilation contributions and the low partial waves have been
studied experimentally. In KN scattering there are I=0 and I=1 channels, and
both are repulsive, with I=1 having the strongest interaction.

These features of KN interactions are correctly predicted
by the quark Born formalism. The
well-determined I=1 KN scattering length of  $\approx -0.33$ fm is predicted to
be about $-0.35$ fm given standard quark model parameters (dashed).
The I=0 scattering
length is predicted to be about $-0.12$ fm, but is unfortunately not yet well
determined experimentally. At higher energies we find that the Born-order
S-wave phase shifts are too ``soft"; the predicted phase shifts are largest at
$P_{cm}\approx 0.5$ GeV, whereas the observed ones appear to grow
monotonically to the maximum experimental momentum of $P_{cm}\approx 0.7$ GeV.
This may be an artifact of our ``soft" Gaussian baryon wavefunctions, since a
good fit is possible given a moderately reduced nucleon length scale (solid).
Higher KN partial
waves are also interesting in that they show clear evidence of spin-orbit
interactions, which we have not yet incorporated in our formalism.

\vskip 0.5cm
\noindent
{\it d) }NN
\vskip 0.2cm

\begin{wrapfigure}{r}{4.0in}
\epsfig{file=h4.epsm,width=4.0in}
\caption{I=0 and I=1 low-energy NN potentials.}
\label{fig 4}
\end{wrapfigure}
Finally, the short-range repulsive core in the NN interaction provides
an important test of any description of
hadron-hadron scattering. We assume that
the repulsive core is dominantly due to the OGE spin-spin hyperfine
interaction, which was a conclusion of resonating group
and variational studies
\cite{hyperf}. We do not expect to reproduce the longer-ranged attraction,
which is variously attributed to spatial wavefunction distortion or meson
exchange; neither effect is present in our formalism at Born order.

When modelled as a potential the short-distance core is typically found to have
a maximum of $\approx 1$ GeV and a range of about 1/2 fm for both NN isospin
states. In our formalism this is a straightforward calculation \cite{BSNN}, and
the two parameters $\alpha_s/m_q^2$ and $\alpha_N$ (baryon
Gaussian width parameter)
are already determined. The
low-energy NN potentials we find are shown in figure 4.

These
core potentials are very similar to the results of resonating
group calculations. Thus, our rather surprising conclusion is that the
cores are dominantly Born-order one-gluon-exchange effects. The
Born-order S-wave NN phase shifts in contrast are quite inaccurate,
since the potentials are so large.
We have used numerical integration of the Schr\"odinger equation to
determine phase shifts which result from the Born-order NN
core potentials, and the resulting
phase shifts closely resemble those of Oka
and Yazaki \cite{hyperf}.

\vskip 0.2cm
\noindent
{\it e) Other channels; vector-vector molecules}
\vskip 0.2cm

In a search for possible molecule states we have investigated many other
meson-meson, meson-baryon and baryon-baryon channels. In nonstrange
S-wave baryons
we found that only the I=0, S=1 and I=1, S=0 $\Delta\Delta$ channels have
attractive cores.
We agree with Maltman \cite{Kim} that the strongest attraction
is in I=0, S=1 $\Delta\Delta$,
although with our parameters
the attraction is not strong enough for binding. We also studied $N_s=1$
and $N_s=2$ octet-octet strange baryon channels \cite{BW},
and find (in agreement
with the resonating group results of
Oka, Shimizu and Yazaki \cite{OSY})
that in $N_s=2$ only I=0, S=0 $\Sigma\Sigma$
has an attractive core. Our study of the coupled
$\Lambda\Lambda-N\Xi-\Sigma\Sigma$ system using the multichannel Schr\"odinger
equation \cite{BW}
sees no evidence for an ``H"
bound state, again in agreement with Oka et al, although a weakly bound
molecular type state may yet exist since we have modelled
only the core forces.
In the strange meson-baryon system we found several channels in the
K$^*$N, K$^*\Delta $ and K$^*\Delta$ systems with
attractive cores, which are not strong enough for binding but may give
threshold enhancements that are normally reported as ``Z$^*$ resonances".

Only in the vector-vector system do we find sufficiently strong interactions
to lead to new molecular bound states.
Bound states form most easily in strongly-coupled systems that are nearly
degenerate, since there are small energy denominators, and one linear
combination of states in a $2\times 2$ Hamiltonian with an off-diagonal
interaction always experiences an attractive interaction. The $qs\bar q \bar s$
vector-vector system is such a case; here Swanson finds an especially large
K$^*\bar{\rm K}^* \leftrightarrow \omega\phi$ matrix element, which in a
multichannel Schr\"odinger formalism described by Dooley, Swanson and Barnes
\cite{DSB} leads to scalar and tensor bound states. These vector-vector
molecules might be identified with the $\theta(1710)$; similar suggestions for
this and other unusual resonances have appeared in the literature
\cite{Kala,Torn,EK}.
The couplings and branching fractions of
the linear combination
$|\theta\rangle = (|{\rm K}^*\bar {\rm K}^*\rangle +
|\omega\phi\rangle )/\sqrt{2}$
which is formed by a large
off-diagonal interaction explain
many of the unusual
properties of the $\theta$.
New predictions of this model
\cite{DSB} include a large branching fraction to
K$\bar {\rm K}\pi\pi$ ($B.F.\approx 35\% $),
an unusual $\phi\pi^0\gamma$ mode ($B.F.\approx 0.3\% $),
$B.F. {\rm K}\bar {\rm K} / B.F. \eta\eta \approx 2$,
a weak $\pi\pi$ mode ($B.F.\sim 5-10\% $),
the $\psi$-hadronic ratio
$(\psi\to\phi\theta) / (\psi\to\omega\theta) \approx 2/\lambda^2$
(just as for
a ${\rm K}\bar {\rm K}$ molecule), and a small
$\gamma\gamma$ width \cite{thetagams}.

\section{Summary and Conclusions}

We have found that a simple perturbative mechanism, one-gluon-exchange followed
by constituent interchange, leads to an accurate description of low energy
hadron-hadron scattering in several annihilation-free channels, given standard
quark-model parameters. In future work it will be interesting to generalize
these techniques through
1) the inclusion of other terms such as spin-orbit (needed
for KN higher partial waves),
2) the use of more realistic baryon
wavefunctions (for KN at higher energies),
3) the incorporation of $q\bar q$
annihilation, which is known to be a very important effect when allowed,
as in K$\bar {\rm K}$
and I=0 $\pi\pi$ scattering (we have completed an initial study of
I=0,1 $\pi\pi$ and I=1/2 K$\pi$ \cite{LGBS}), and
4) the incorporation of t-channel pion exchange forces, which dominates the
long range force when allowed.
With these improvements we expect
that it will be possible to predict phase shifts and bound states accurately in
other meson-meson and meson-baryon channels, which will be accessible to
experiments at hadron facilities such as CEBAF, DAPHNE, KAON, KEK and LEAR. Our
comparisons with experiment suggest that better determinations of some
scattering amplitudes, such as I=2 $\pi\pi$ (above 1.5 GeV), I=3/2 K$\pi$ and
especially I=0 KN, would be very useful contributions to the study of soft
hadron scattering.

\section{Acknowledgements}

It is a pleasure to acknowledge the efforts of the organisers of this meeting,
in particular C.Carlson and F.Gross. The advice and
essential contributions of my collaborators S.Capstick,
K.Dooley, G.Grondin, N.Isgur, M.D.Kovarik, Z.P.Li, E.S.Swanson and
J.Weinstein
are also gratefully acknowledged. This research was sponsored in
part by the United States Department of Energy under contract
DE-AC05-840R21400, managed by
Martin Marietta Energy Systems, Inc.

\end{document}